\documentclass[%
 reprint,
 amsmath,amssymb,
 aps,
]{revtex4-2}

\usepackage{graphicx}
\usepackage{dcolumn}
\usepackage{bm}
\usepackage[colorlinks=true,bookmarks=false,citecolor=blue,urlcolor=blue]{hyperref}
\usepackage{comment}
\usepackage{multirow}

\begin{document}

\preprint{APS/123-QED}

\title{Full transmission of vectorial waves through 3D multiple-scattering media}

\author{Ho-Chun Lin}
\author{Chia Wei Hsu}%
\email{cwhsu@usc.edu}
\affiliation{%
Ming Hsieh Department of Electrical and Computer Engineering, University of Southern California, Los Angeles, CA 90089, USA
}

\begin{abstract}
A striking prediction from the random matrix theory in mesoscopic physics is the existence of ``open channels'': waves that can use multipath interference to achieve perfect transmission across an opaque disordered medium even in the multiple-scattering regime. Realization of such open channels requires a coherent control of the complete incident wavefront. To date, the open channels have only been demonstrated in scalar two-dimensional (2D) structures, both experimentally and with numerical studies. Here, we utilize a recently proposed ``augmented partial factorization'' full-wave simulation method to compute the scattering matrix from 3D vectorial Maxwell's equations and demonstrate the existence of open channels in 3D disordered media. We examine the spatial profile of such open channels, demonstrate the existence of a bimodal transmission eigenvalue distribution with full control, and study the effects of incomplete polarization control and of a finite illumination area. This study confirms the validity of the random matrix theory in vectorial systems. The simulation framework provides full access to the complex multi-channel wave transport in 3D disordered systems, filling the gap left by experimental capabilities.
\end{abstract}

\maketitle

\noindent
{\bf Introduction.} Light propagation through opaque disordered media presents both challenges and opportunities due to the interplay between multiple scattering events~\cite{Mosk2012_NP, 2013_Wiersma_nphoton_review,2017_Stefan_RMP,2022_Cao_NP}. These scattering events can lead to counterintuitive phenomena not captured by classical diffusion theory, like coherent backscattering~\cite{2007_Akkermans_book}, Anderson localization~\cite{2009_Lagendijk_PT}, and coherent enhancement of remission~\cite{2022_Bender_PNAS}. A particular striking phenomenon is the existence of ``open channels'', where customized incident wavefronts propagate through opaque media with near-perfect transmission~\cite{1984_Dorokhov_SSC,1988_Mello_AnnPhys,1994_Nazarov_PRL,1997_Beenakker_RMP}. The open channel arises from the constructive interference of multiple scattered waves and is intrinsically related to the long-range correlations between them~\cite{1988_Mello_PRL,2017_Hsu_nphys,Bender2020_PRL}. Its near-perfect transmission allows light to penetrate through opaque media and deliver energy efficiently, paving the way for many potential applications such as optical communication~\cite{2018_Tzang_NP,2021_Nie_OE}, deep optical imaging~\cite{2020_Yoon_NRP,2022_Gigan_JoPP}, and encryption~\cite{2021_Ruan_NC}.

The existence of open channels was first predicted by the random matrix theory (RMT) of Dorokhov-Mello-Pereyra-Kumar (DMPK)~\cite{1984_Dorokhov_SSC,1988_Mello_AnnPhys,1994_Nazarov_PRL} in the context of electron transport. Despite this prediction, such open channels have never been observed in three dimensions (3D) either in experiments or simulations. To date, experimental~\cite{2014_Gerardin_PRL,2021_Meer_arXiv,2022_Horodynski_Nature} and numerical~\cite{2011_Choi_PRB,2013_Goetschy_PRL,2015_Liew_OE,2015_Hsu_PRL,2020_Liu_Optica,2022_Valantinas_arXiv} observations of open channels all consider 2D systems. Experiments in 3D observed an enhanced transmission~\cite{2013_Yu_PRL,2014_Popoff_PRL,2017_Hsu_nphys, 2020_Bosch_phD_thesis}, which remained much lower than unity because of the incomplete wavefront control and collection due to finite-area illumination and the finite numerical aperture of objective lenses~\cite{2013_Goetschy_PRL,2014_Popoff_PRL}. Numerical simulations in 3D, meanwhile, require immense computing resources given the large system size and the large number of input modes. One can speed up the simulations by constructing preconditioners for iterative methods~\cite{Osnabrugge_2016_JCP, 2019_Vettenburg_OE, 2022_vettenburg_arXiv} or using hardware acceleration~\cite{2017_CELES, 2018_Pattelli_Optica, 2022_Valantinas_arXiv, 2023_Yamilov_NPhys}, but these methods require one simulation for every input of interest, so it would take a very long time to build the full transmission matrix required to study open channels. Thus, the existence of these perfectly transmitting open channels in 3D vectorial systems remains unverified, and their unique properties have not been studied explicitly.

We recently introduced the numerical method of ``augmented partial factorization'' (APF)~\cite{2022_Lin_NCS}, which efficiently solves multi-input electromagnetic scattering problems in a single step. The work in Ref.~\cite{2022_Lin_NCS} considered 2D scalar Maxwell's equations without parallelization. Here, we implement a parallelized version of the APF method for 3D vectorial Maxwell's equations, and use it to perform numerical simulations that demonstrate the existence of 3D open channels for the first time. We also study the effects of incomplete polarization control and finite illumination area. 

\begin{table*}[ht!]
\centering
\caption{\label{Tab:computing_resource}{\bf Computing costs using the APF method and a conventional direct method.} There are 1482 input modes in the transmission matrix {\bf{t}}, 2964 input modes in the scattering matrix {\bf{S}}.}
\setlength{\extrarowheight}{-2pt} 
\begin{tabular}{cccccc}
\hline

                & {Memory} & {Disk} & \multicolumn{3}{c}{Computing time (hour)} \\
                & usage (GiB)                                               & usage (GiB)                                  & One input     & Single {\bf t}        & Full {\bf S}            \\
\hline
APF method             & 205                                            & 0                                   & 2.36        & 2.36              & 2.36              \\
{\multirow{2}{*}{Conventional direct method}} \ \ Storing LU factors in RAM & $\sim  600$                                           & 0                                   & N/A                   & N/A             & N/A   \\
\ \ \ \ \ \ \ \ \ \ \ \ \ \ \ \ \ \ \ \ \ \ \ \ \ \ \ \ \ \ \ \ \ \ \ \ \ \ \ \ \ \  \ \ \ \ \ \ \ \ Storing  LU factors in disk & 167                                            & 518                                 & 2.44      &$\sim  430$            & $\sim  850$   \\
\hline
\end{tabular}
  \label{tab:shape-functions}
\end{table*}

{\bf Transmission matrix computation with APF.} The open channel has a near-perfect transmission. The incident wavefront that maximizes the transmission is the right singular vector of the transmission matrix {\bf t} with the largest singular value $\sigma_n$~\cite{2017_Stefan_RMP}. The transmission matrix {\bf t}~\cite{2010_Popoff_PRL,Mosk2012_NP,2017_Stefan_RMP} relates the incident wavefront ${\boldsymbol \beta} = [\beta_1,...,\beta_q,...,\beta_{{M_{\rm{in}}}}]$ written as a vector to the transmitted wavefront ${\boldsymbol \alpha} = [\alpha_1,...,\alpha_q,...,\alpha_{{M_{\rm{out}}}}]$,
\begin{equation}
{{\alpha}_p} = \sum_{q=1}^{M_{\rm in}} {{t_{p q}}{{\beta}_q}},
 \label{eq:t_eq}
\end{equation} 
where $\beta_q$ and $\alpha_p$ represent the complex amplitude of the
$q$-th and $p$-th propagating modes on the input and transmitted side; $M_{\rm in}$ and $M_{\rm out}$ are number of such modes. 
To find the most open channel and its incident wavefront ${\boldsymbol \beta}^{\rm open}$, we proceed by computing the transmission matrix {\bf t} for all of its $M_{\rm in}$ distinct inputs.

To compute the transmission matrix {\bf t}, here we numerically solve the 3D vectorial Maxwell's equations that govern light scattering and propagation inside the disordered medium. The discretized Maxwell’s equations in the frequency domain can be written as a system of linear equations ${\bf A}{\bf x}_q = {\bf b}_q$, where ${\bf A} = - (\omega/c)^2 {\bar{\bar \varepsilon}} _{\rm{r}}(\omega,{\bf{r}}) + \nabla  \times \nabla  \times$ is the electric-field operator at frequency $\omega$, $c$ is the speed of light, and ${\overline{\overline{\varepsilon}}}_{\rm r}(\omega,{\bf{r}})$ is the spatial profile of the relative permittivity tensor that describes the disordered medium. The column vector ${\bf b}_q({\bf{r}})$  on the right-hand side is a source profile that generates the $q$-th incident wavefront, and the column vector ${\bf x}_q({\bf{r}})$ contains the resulting electric field profile ${\bf E}_q({\bf{r}}) = (E_x, E_y, E_z)$. Given the $q$-th solution ${\bf x}_q  = {\bf A}^{-1}{\bf b}_q$ of the system of linear equations, one can project it onto each of the $M_{\rm out}$ transmitted modes to obtain the transmitted wavefront as ${\boldsymbol \alpha} = {\bf C}{\bf x}_q$, where the rows of matrix ${\bf C} = [{\bf c}_1; ...; {\bf c}_{M_{\rm out}}]$ contains the projection profiles. The transmission matrix {\bf t} can then be computed through 
\begin{equation}
{\bf t} = {\bf C} {\bf A}^{-1}{\bf B},
\label{eq:t_CA-1B}
\end{equation} 
where the input matrix ${\bf B} = [{\bf b}_1, ..., {\bf b}_{M_{\rm in}}]$ contains the $M_{\rm in}$ source profiles. To efficiently evaluate Eq.~\eqref{eq:t_CA-1B}, we adopt the recently introduced APF method~\cite{2022_Lin_NCS}. With APF, we build an augmented matrix ${\bf{K}} = [{\bf A}, {\bf B}; {\bf C}, {\bf{0}}]$ and perform a partial factorization on it, which yields the Schur complement $-{\bf C}{\bf A}^{-1}{\bf B} = -{\bf t}$ of {\bf{K}}. This directly provides the entire transmission matrix ${\bf t}$ without computing the electric field distribution ${\bf x}_q = {\bf A}^{-1}{\bf b}_q$ everywhere in space, without looping over the $M_{\rm in}$ distinct inputs, and without storing the factors of {\bf A}.

Here, we discretize the 3D vectorial electric-field operator ${\bf A}$ on the Yee grid~\cite{Yee1966_TAP,1994_Chew_JAP}, where different components of the ${\bf E}$ and ${\bf H}$ fields are staggered in space to achieve second-order accuracy. Perfectly matched layers (PMLs)~\cite{2022_Pled_ACME} are implemented to realize outgoing boundaries. In addition, we perform subpixel smoothing on the refractive index profile~\cite{2006_Farjadpour_OL,2009_Oskooi_OL}, resulting in an anisotropic discretized ${{\bar {\bar {\varepsilon}} }_{{\rm r}}}$ tensor. We also derive and implement the source matrix ${\bf B}$ and projection matrix ${\bf C}$ that generates and projects 3D plane waves on the Yee grid for both polarizations. We use the MUMPS linear solver~\cite{2019_Amestoy_ACM} with single-precision arithmetic and the METIS ordering~\cite{1998_Karypis_SIAM} to perform the partial factorization with share-memory and distributed-memory parallelization. All computations are carried out on an AMD epyc-7513 node with 32 cores (64 threads) and 256 GiB of RAM with a Western Digital 12TB HDD.

\begin{figure}[h!]
\centering\includegraphics[width=0.45\textwidth]{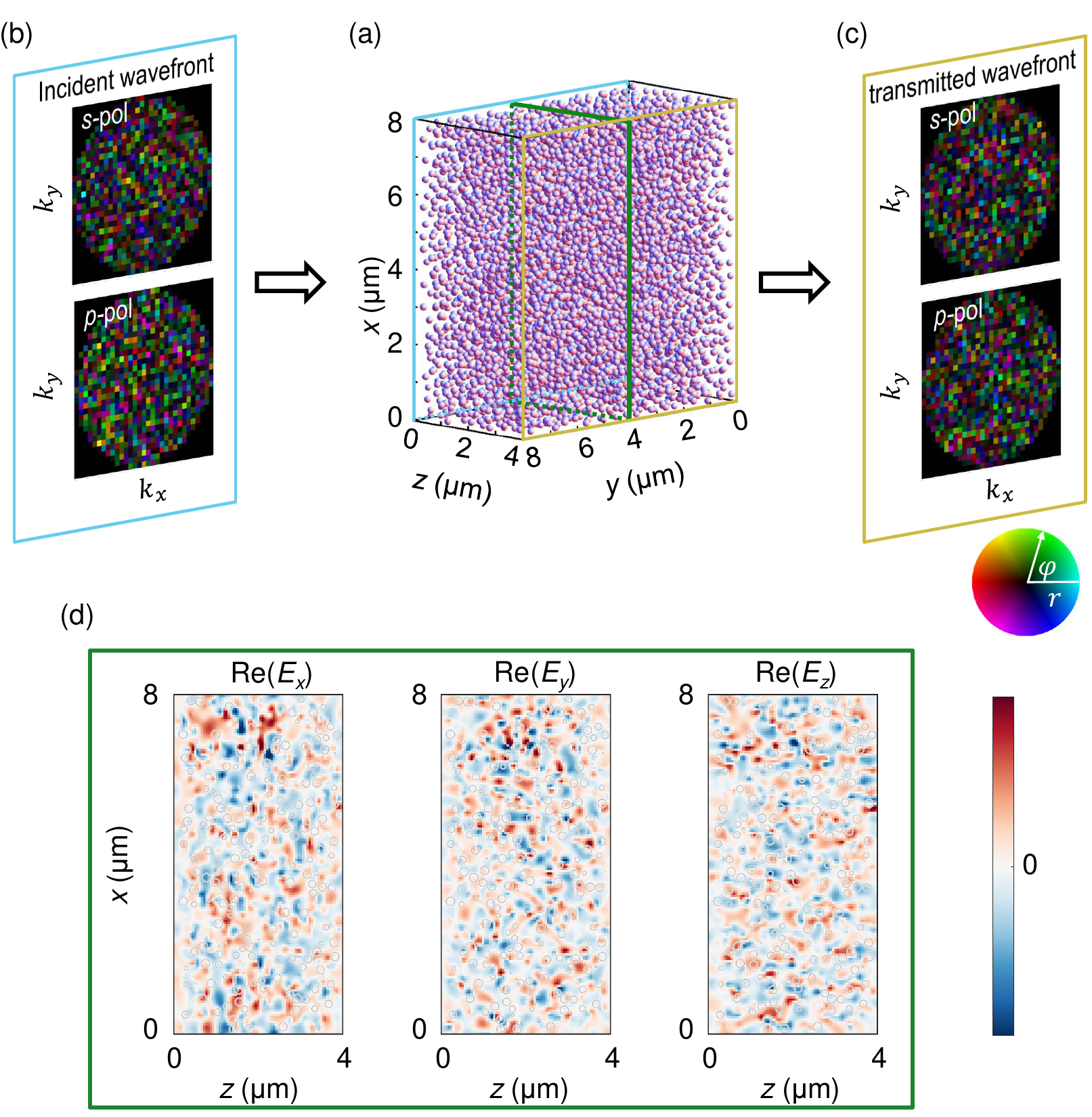}
\caption{\label{fig:schematics}{\bf 3D vectorial open channel.}
(a) The disordered medium considered in the simulation, consisting of 7,200 TiO$_2$ spheres with diameter 200~nm, at wavelength $\lambda_0 = 532$ nm. The average transmission is 0.097. (b) The phase $(\varphi)$ and amplitude $(r)$ of the open-channel incident wavefront $\boldsymbol{\beta}^{\rm open}$ at $z = 0$ (cyan plane) in the momentum space $(k_x,k_y)$ for both polarizations, obtained from the transmission matrix, which is computed by solving the 3D vectorial Maxwell's equations with APF. (c) The corresponding transmitted wavefront $\boldsymbol{\alpha}^{{\rm open}}$ at $z = L$ (yellow plane). The transmission is around unity. (d) The longitudinal field profiles $E_x$, $E_y$, and $E_z$ of the open channel at $y = 4$ \textmu m (green plane). Circles indicate the cross sections of the TiO$_2$ spheres.
}
\end{figure}

We consider 3D disordered media with thickness $L = 4$ \textmu m in $z$ and width $W = 8$ \textmu m in $x$ and $y$, at free-space wavelength $\lambda_0 = 532$ nm. The disordered medium consists of 7,200 spherical TiO$_{\rm{2}}$ nonoparticles in air~\cite{1994_Boer_PRL, 2005_Lodahl_PRL}, with refractive index 2.54~\cite{2016_Bodurov_NN}, diameter 200 nm, and volume fraction 0.12 (Fig~\ref{fig:schematics}(a)). Periodic boundary conditions are applied in $x$ and $y$ to mimic an infinitely wide slab, and PMLs are used in $z$ for outgoing boundaries. We discretize with grid size $\Delta x = \lambda_0/10$, leading to 2.7 million voxels. Counting both polarizations, there are 1482 propagating modes on the input side ($z<0$) of the disordered medium, and similarly on the transmitted size ($z>L$). For each disordered medium, we compute the transmission matrices from both sides (which can be used for ensemble average). With APF, we directly compute the full scattering matrix {\bf S} of the system, which contains both transmission matrices.
The ensemble-averaged transmission is 0.097. 

As shown in the first row of Table~\ref{Tab:computing_resource}, it takes 2.36 hours and 205 GiB of memory to compute the full scattering matrix {\bf{S}} (containing both transmission matrices) of this system with APF. The computing time is independent of the number of inputs.

As a comparison, we also implement a conventional direct method to compute {\bf t}. The conventional direct method also computes the transmission matrix through Eq.~\eqref{eq:t_CA-1B}, but it does so by factorizing the matrix {\bf A} into upper-lower (LU) factors, solving for ${\bf A}^{-1}{\bf b}_q$ for the $q$-th input using the LU factors, projecting it with ${\bf C}$, and looping $q$ over the $M_{\rm in}$ inputs. Storing the LU factors would take up more RAM than the machine has, so we have to store them in the hard drive, which further slows down the solution. While reusing the LU factors for the $M_{\rm in}$ inputs substantially speeds up the solution, the total computing time still grows linearly with the number of inputs. Resultingly, computing the full scattering matrix {\bf{S}} would take an estimated 850 hours (more than one month), in stark contrast with the 2.36 hours of APF (Table~\ref{Tab:computing_resource}). Doing so is challenging even for one disorder realization, and the ensemble average would be practically infeasible.

{\bf 3D open channel.}
After computing the transmission matrix with APF, we perform singular value decomposition on it. To validate our calculations, supplementary Fig.~1 demonstrates the unitarity of all right singular vectors, confirming that their sum of transmission and reflections nearly equals unity (slight deviations stem from the error caused by PMLs.) We then obtain the most open channel and its incident wavefront $\boldsymbol{\beta}^{\rm open}$ and the transmission is $\sigma_1^2 = 1.00$, substantially above the average transmission of 0.097. Figure~\ref{fig:schematics}(b) shows the two polarization components of $\boldsymbol{\beta}^{\rm open}$ for all incident angles. Its resulting transmitted wavefront $\boldsymbol{\alpha}^{{\rm open}} = {\bf t} \boldsymbol{\beta}^{\rm open}$ is shown in Fig.~\ref{fig:schematics}(c). Given $\boldsymbol{\beta}^{\rm open}$, we also compute the field profile of the open channel inside the disordered medium using a conventional direct solver, shown in Fig.~\ref{fig:schematics}(d) for the $E_x$, $E_y$, and $E_z$ components at $y = 4$ \textmu m. Supplementary Fig.~2 further shows the $x$-$y$-integrated intensity as a function of depth.

{\bf Bimodal transmission eigenvalue distribution.}
Going beyond the open channel, here we further examine the full DMPK theory, which has not been validated for 3D vectorial waves before. In the diffusive regime ($L \gg l$, where $l$ is the transport mean free path), the DMPK theory predicts that the eigenvalues $\tau_n$ of the Hermitian matrix ${{{\bf t}^\dag }{\bf t}}$ has a bimodal probability distribution~\cite{1984_Dorokhov_SSC,1986_Imry_Europhys_Lett,1992_Pendry_PRSL},
\begin{equation}
\label{eq:bimodal_distr}
{p_{{{ \bf t}^\dag }{ \bf t}}}\left( \tau \right) = \frac{{\bar \tau}}{{2\tau\sqrt {1 - \tau} }},
\end{equation}
where ${\bar \tau} = \langle \tau_n \rangle $ is the average transmission. The transmission eigenvalues equal the square of the singular values of ${\bf t}$, $\tau_n = \sigma_n^2$. Eq.~\eqref{eq:bimodal_distr} shows a fraction ${\bar \tau}$ of all transmission eigenvalues are ``open'' with order-unity transmission, and they are responsible for the overall transmission of the system, with the other eigenvalues being negligible small (``closed'').

\begin{figure}[t!]
\centering\includegraphics[width=0.45\textwidth]{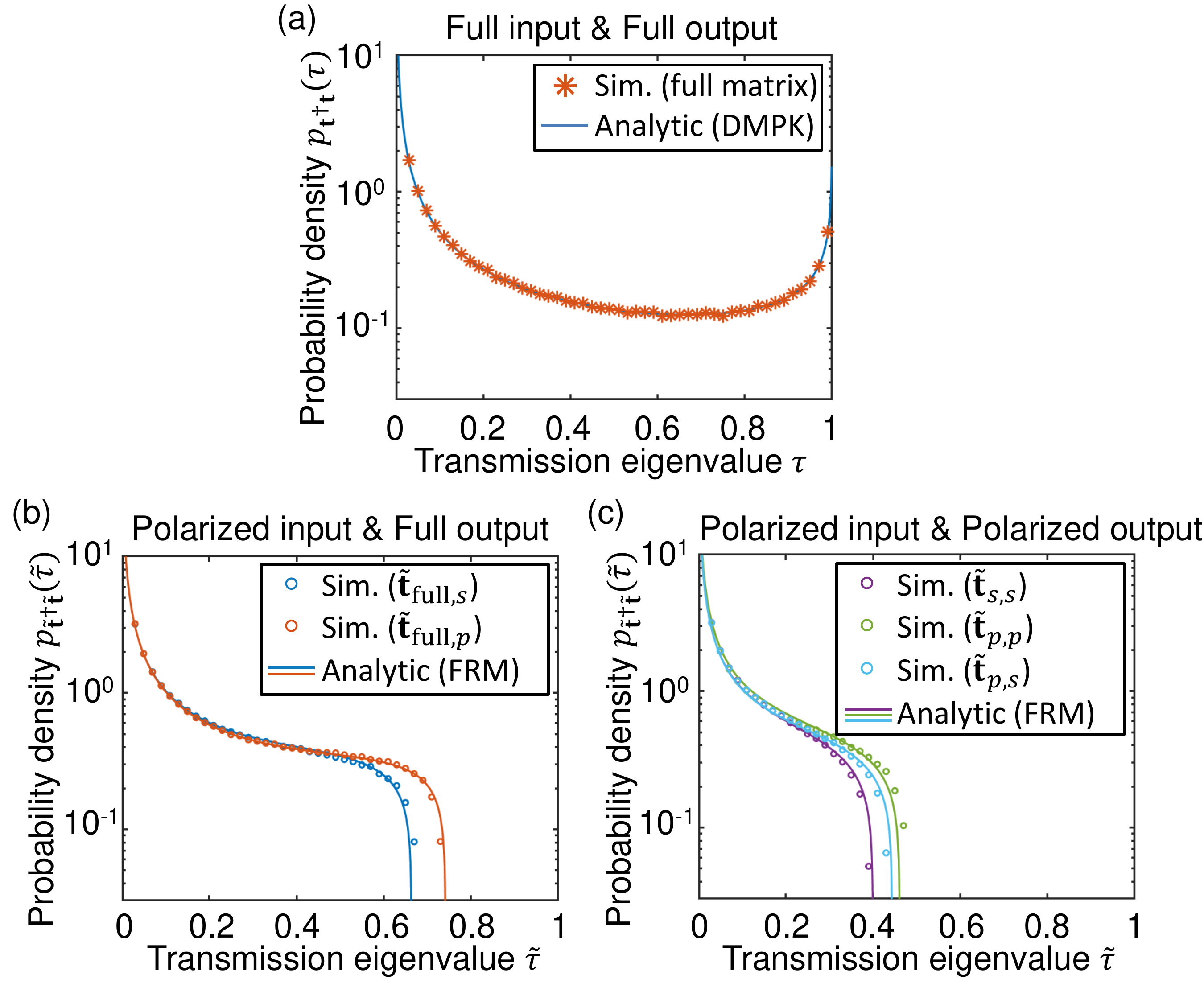}
\caption{\label{fig:complete_and_incomplete_control}{\bf Transmission eigenvalue distributions and polarization effects.} (a) The probability density of the eigenvalues of ${{{\bf t}^\dag }{\bf t}}$. (b) Eigenvalue distribution when only one polarization is used for the incident wave. (c) Eigenvalue distribution when only one polarization is used for both the incident wave and the detection. Symbols are from the APF simulations, computed from 200 realizations of ${\bf t}$. Solid lines are analytic predictions from the DMPK random matrix theory and the filtered random matrix (FRM) theory.}
\end{figure}

To capture the statistical distribution, we compute 200 ${\bf t}$ matrices for different disorder realizations and obtain ${p_{{{\bf t}^\dag }{\bf t}}}\left( \tau \right)$. The numerical ${p_{{{\bf t}^\dag }{\bf t}}}\left( \tau \right)$ is shown in asterisks in Fig.~\ref{fig:complete_and_incomplete_control}(a), which agrees strikingly well with the analytic prediction from Eq.~\eqref{eq:bimodal_distr}. To our knowledge, this is the first time the DMPK distribution is confirmed in a 3D vectorial system. 

{\bf Incomplete polarization control.} 
Experimental observation of the open channels in 3D is difficult because of the unavoidable incomplete control arising from not exciting and/or detecting both polarizations, a finite numerical aperture, and having a finite illumination area~\cite{2013_Goetschy_PRL, 2013_Yu_PRL,2014_Popoff_PRL}. Here, we numerically study these effects.

To investigate the polarization effects, we compute the eigenvalues $\tilde \tau$ of matrix $\Tilde{{\bf t}}^{\dag}\Tilde{{\bf t}}$, where $\Tilde{{\bf t}}$ is a filtered transmission matrix~\cite{2013_Goetschy_PRL,2014_Popoff_PRL} that contains only one of the two polarization modes in the input and/or output. Figure~\ref{fig:complete_and_incomplete_control}(b)--(c) shows the eigenvalue distributions with the full output and one polarization in the input, and with one polarization in both the input and the output. When we do not fully control all available modes, the peak at $\tau \approx 1$ corresponding to the open channels disappears. To compare with the random-matrix prediction, here we also adopt the filtered random matrix (FRM) theory to compute the analytic eigenvalue distribution, accounting for the fact that the two polarizations have unequal average transmission (Supplementary Sec.~3).

\begin{figure}[t!]
\centering\includegraphics[width=0.43\textwidth]{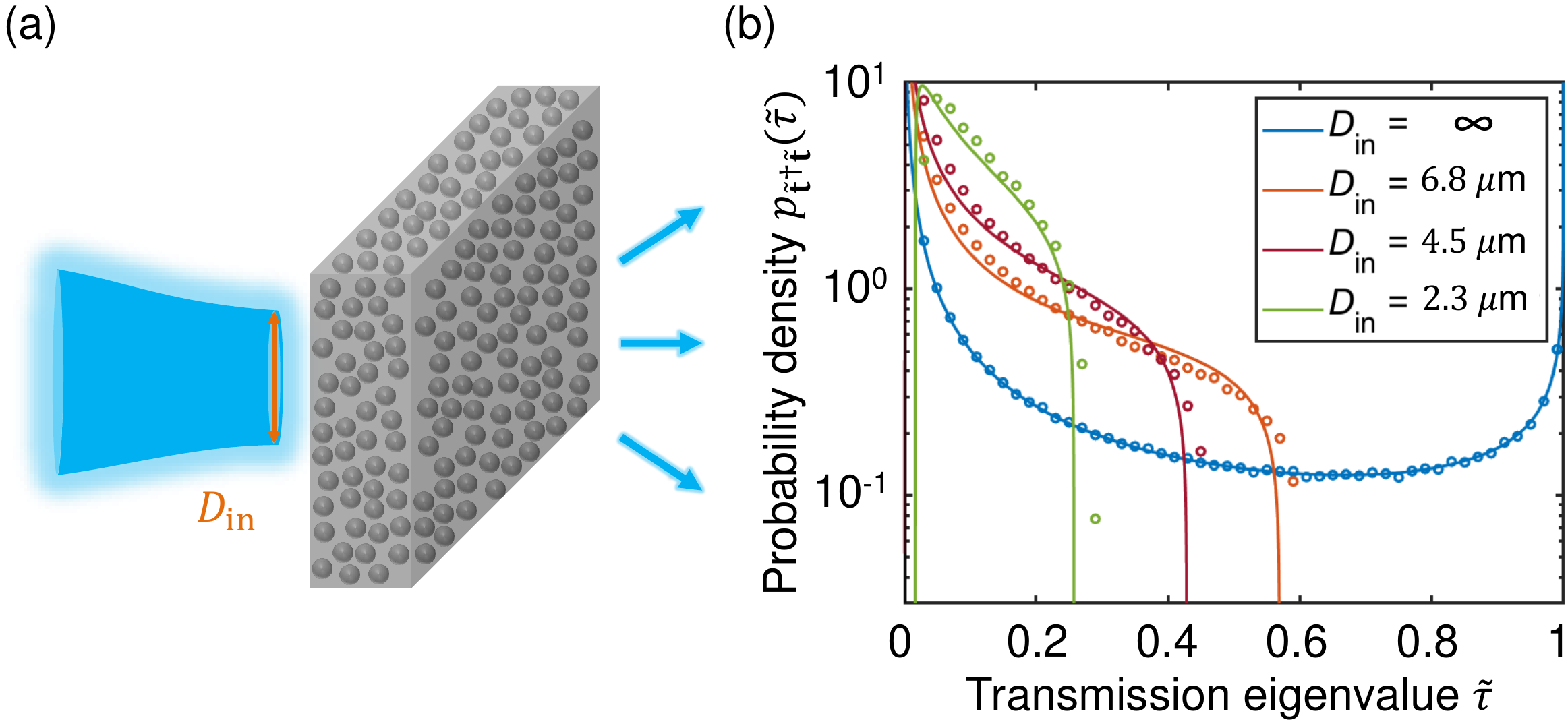}
\caption{\label{fig:finite_illumination}{\bf Effects of a finite illumination area.} (a) Schematic of a finite-area illumination. (b) The transmission eigenvalue distributions with different illumination diameters $D_{\rm{in}}$. Symbols are from simulations, and solid lines are FRM predictions. }
\end{figure}

{\bf Finite-area illumination.}
In an experiment, a finite illumination area at $z=0$ (Fig.~\ref{fig:finite_illumination}(a)) corresponds to having a finite angular spacing in the momentum space~\cite{2017_Hsu_nphys} Therefore, to model a finite-area illumination, we group the input modes into clusters of ``macropixels'' in the momentum space, with each macropixel comprising $n_0^2$ adjacent angles, spanning a size of $q = n_0(2\pi/W)$ in the momentum space. This results in a ${\rm sinc}^2(qx/2){\rm sinc}^2(qy/2)$ illumination profile with diameter $D_{\rm{in}} = 3\pi/q$~\cite{2017_Hsu_nphys}. As shown in Fig.~\ref{fig:finite_illumination}(b), the open-channel peak immediately disappears when a finite illumination area is adopted. This explains the challenge of observing these open channels in an open geometry. We also compare the numerical results to the theoretical predictions from FRM (Supplementary Sec.~4).

{\bf Discussion.} In summary, we have implemented a parallelized APF method for 3D vectorial Maxwell's equations and used it to demonstrate the existence of open channels, validate the DMPK bimodal distribution, and study the effects of incomplete polarization control and finite-area illumination. We have made our code open-source~\cite{MESTI_jl_GitHub}, which one can readily modify to perform multi-input full-wave simulations of wave propagation in complex 3D systems. Beyond open channels, many studies on wavefront shaping in disordered systems currently rely on scalar 2D simulations, such as transverse localization~\cite{2019_Yılmaz_NPhotonics}, branched flow~\cite{2019_Brandstötter_PNAS}, scattering invariant mode~\cite{2021_Pai_NPhotonics}, tractor beams~\cite{Horodynski_PRA_2023}, dwell-time operator~\cite{2019_Durand_PRL}, energy-shift operator~\cite{Hüpf_PRA_2023}, and the generalized Wigner-Smith operator~\cite{Ambichl_PRL_2017}, as well as conductance computation for correlated disordered systems~\cite{2017_Froufe-Perez_PNAS}. Our code enables these and other studies to be carried out in full 3D vectorial form. It can also be used for the design of multi-channel nanophotonic structures such as wide-field-of-view metalenses~\cite{2024_Li_Optica} and multiplexers~\cite{2020_Hougne_PRApplied, 2022_Plückelmann_Nanophotonics}. 

{\bf Acknowledgments:} Computing resources are provided by the Center for Advanced Research Computing at the University of Southern California. 
{\bf Funding:} National Science Foundation (ECCS2146021), Sony Research Award Program, and DARPA.
{\bf Disclosures:} The authors declare no conflicts of interest. {\bf Data availability:}
All data needed to evaluate the conclusions in
this study are presented in the paper and supplemental document. The code is available on GitHub~\cite{MESTI_jl_GitHub}.

\bibliography{main}

\end{document}


\title{Full transmission of vectorial waves through 3D multiple-scattering media: supplemental document}

\author{Ho-Chun Lin}
\author{Chia Wei Hsu}%
\email{cwhsu@usc.edu}
\affiliation{%
Ming Hsieh Department of Electrical and Computer Engineering, University of Southern California, Los Angeles, California 90089, USA
}%

%

\maketitle
\section{Unitarity test with the single-precision}
\begin{figure}[h!]
\centering\includegraphics[width=0.5\textwidth]{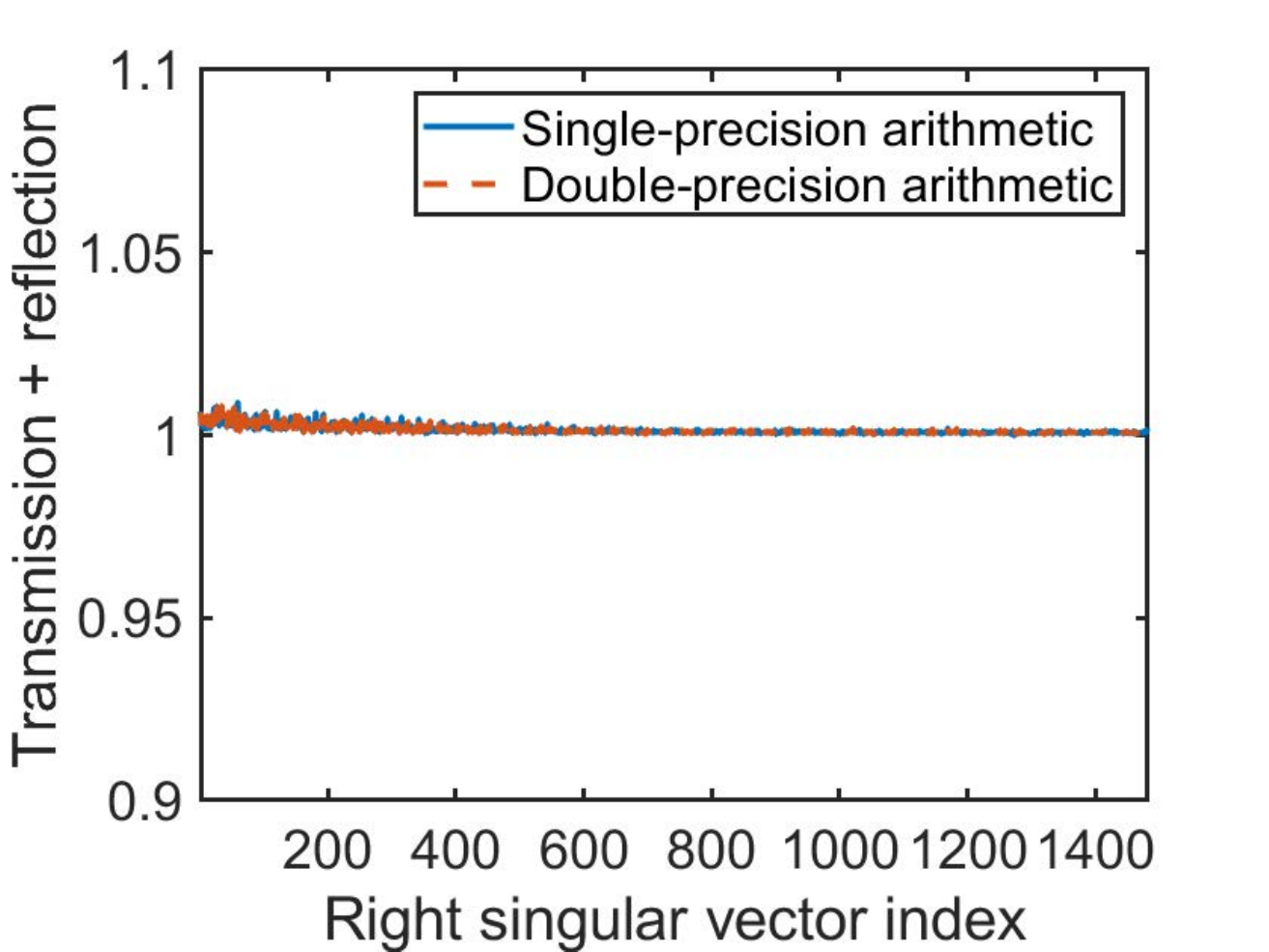}
\caption{\label{fig:untarity}{\bf Unitarity test with the two arithmetics.} Unitarity test for right singular vectors of the transmission matrix {{\bf t}} computed using the single-precision arithmetic and the double-precision arithmetic. Both results conserve unitarity, with the sum of transmission and reflection  approaching 1.}
\end{figure}

In the main text, all the results come from the single-precision arithmetic from the MUMPS linear solver. Here, we validate our calculations in transmission matrix {\bf t} with the single-precision arithmetic by comparing it with {\bf t} from the double-precision arithmetic. We compute {\bf t} for the same system described in the main text with the single-precision arithmetic and the double-precision arithmetic. We then perform singular value decomposition on the transmission matrix {\bf t} and obtain their right singular vectors representing the wavefronts for different transmissions. Since our system is in the absence of absorption or gain, we can test the unitarity for these right singular vectors: for each right singular vector, the sum of total reflection and total transmission should be unity. As shown in Fig.~\ref{fig:untarity}, the right singular vectors from both arithmetics conserve unitarity. It validates that single-precision arithmetic has a negligible difference compared to double-precision arithmetic and provides accurate enough results for the purposes of this work. We note that the right singular vectors show the sum of transmission and reflection between 1 and 1.01. This is due to the limitations of perfectly matched layers (PMLs), primarily caused by reflection errors from large angle modes.

\section{$x$-$y$-integrated intensity inside the disordered medium for three distinct incident wavefronts}
\begin{figure}[t!]
\centering\includegraphics[width=0.5\textwidth]{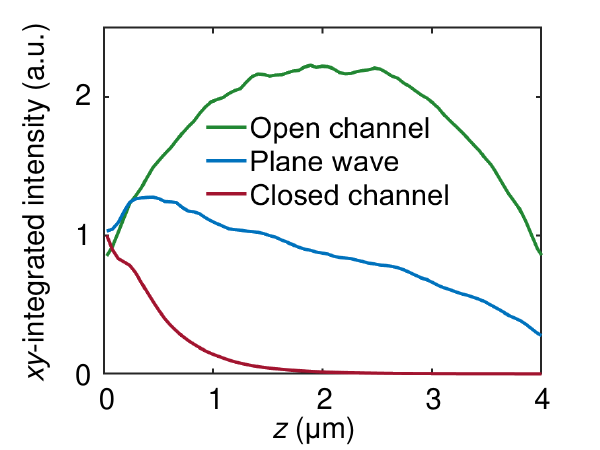}
\caption{\label{fig:xy_integrated_intensity}{\bf $xy$-integrated intensity.} $xy$-integrated intensity as a function of the depth in the $z$ direction for a normal incident plane wave, open channel, and closed channel, respectively. The intensity is normalized relative to that of a closed channel on the incident side.}
\end{figure}

We show the field profile of the most open channel inside the disordered medium in the main text. To further highlight the distinctive nature of open channels, we compute the field profiles for three incident wavefronts injected into the disordered medium: the wavefront for the most open channel, the normal incident mode, and the wavefront for the most closed channel, which corresponds to the right singular vector of transmission matrix {\bf t} with the smallest singular value. Figure.~\ref{fig:xy_integrated_intensity} plots the $x$-$y$-integrated intensity as a function depth $z$ in the disordered medium for these three distinct incident wavefronts. The plane wave input shows a clear linear decay as it propagates into the disordered medium, and the most closed channel input demonstrates an exponential decay. In contrast to the other two inputs, the intensity of the most open channel reaches its peak in the middle of the disordered medium~\cite{2015_Davy_NC,2016_Ojambati_OE}, which is like a Fabry-Perot cavity at the resonance condition due to constructive interference. These observations align well with previous two-dimensional scalar wave study~\cite{2011_Choi_PRB}, and it is the first time to demonstrate their intensity profiles in the 3D system.

\section{Eigenvalue distribution in incomplete polarization control}
In the main text, we consider the effects of incomplete polarization control, which is only exciting and/or detecting modes in certain polarizations. Incomplete polarization control can be encapsulated in an $N_{{\rm out}} \times N_{{\rm in}}$ filtered transmission matrix $\tilde{{\bf t}}$ from an $M_{{\rm out}} \times M_{{\rm in}}$ full transmission matrix ${{\bf t}}$.
$M_{\rm in}$ and $M_{\rm out}$ are the total number of full input modes and output modes. $N_{\rm{in}}$ and $N_{\rm{out}}$ are the number of excitable input and detectable output modes. The filtered transmission matrix $\tilde{{\bf t}}$ contains only one of the two polarization modes in the input and/or output. In addition to numerically computing the transmission eigenvalues $\tilde{{\tau}}$ and their distribution $p(\tilde{{\tau}})$ from $\tilde{{\bf{t}}}^\dagger \tilde{{\bf{t}}}$, one can adopt the equations from filtered random matrix (FRM) theory~\cite{2013_Goetschy_PRL} to analytically predict the eigenvalue distribution.

If the two polarizations are fully mixed and are statistically equivalent, the FRM theory can formulate the analytic form of $p(\tilde{{\tau}})$ based on the fractions of controllable modes $N_{\rm{in}}/M_{\rm{in}}$ and $N_{\rm{out}}/M_{\rm{out}}$. However, here we find the two polarizations to have unequal average transmission: the average total transmission is 0.089 when averaged among $s$-polarized inputs, but the same number becomes 0.104 when averaged among $p$-polarized inputs.

To compensate for the average total transmission difference, motivated by the treatment of angular filtering in Ref.~\cite{2013_Goetschy_PRL}, we additionally use empirical renormalization factors to match the average total transmission difference from $s$-polarized and $p$-polarized inputs/outputs. The renormalization factor is the ratio of the average transmission  $\bar{\tilde{\tau}} = \langle \tilde{\tau} \rangle $ from the filtered transmission matrix $\tilde{\textbf{t}}$ to the average transmission $\bar{{\tau}} = \langle {\tau} \rangle $ from the full transmission matrix \textbf{t}. For example, the renormalization factor for the blue curve (only $s$-polarization input wave and full output) in Fig. 2(b) of the main text is $\bar{\tilde{\tau}}_{{\rm full},s}/\bar{{\tau}}$, where $\bar{\tilde{\tau}}_{{\rm full},s}$ is the average transmission from the filtered transmission matrix ${\tilde{\bf{t}}}_{{\rm full},s}$ with only $s$-polarized input and both polarized output. The renormalization factor for the purple curve (only $s$-polarization input wave and $s$-polarization output wave) in Fig. 2(c) of the main text is ($\bar{\tilde{\tau}}_{{\rm full},s}/\bar{{\tau}}$)(2$\bar{\tilde{\tau}}_{s, {\rm full}}/\bar{{\tau}}$), where $\bar{\tilde{\tau}}_{s,{\rm full}}$ is the average transmission from the filtered transmission matrix ${\tilde{\bf{t}}}_{s, {\rm full}}$ with both polarized input and only $s$-polarized output. The renormalization factor is used to compensate for the average transmission of the filtered transmission matrix $\tilde{\textbf{t}}$. With the renormalization factor and the fractions of controllable modes, we compute the analytic eigenvalue distribution as shown in Fig.~2(b)--(c) in the main text. 

\section{Eigenvalue distribution in finite-area illumination}
In the main text, we group the input modes into clusters of ``macropixels'' in the momentum space to model a finite-area illumination. Here, we define modes in the momentum space and then discuss how to numerically and theoretically obtain the transmission eigenvalues under this finite-area illumination effect. 

Since periodic boundary conditions are applied in this work along $x$-direction and $y$-direction with the transverse width $W$, the mode profile in the transverse direction is $e^{ik_xx+ik_y y}$ with the transverse momentum $k_x$ and $k_y$. The full mode profile $\bf{E}$ should satisfy ${\bf E}(x,y,z) = {\bf E}(x\pm W,y\pm W,z)$, so the transverse momentums for the mode ($\alpha, \beta$) are ($k^{(\alpha)}_x = 2\pi\alpha/W$, $k^{(\beta)}_y = 2\pi\beta/W$). Each mode spans a size of $2\pi/W$ in the momentum space and represents an incident plane wave for different incident angles with infinite illumination area. Intuitively, similar to the Fourier transform, we can decrease the illumination area by incorporating more momenta into the incident waves. So to model the finite-area illumination, instead of using a single mode, we group these input modes into clusters of “macropixels” in the momentum space with $n_0 \times n_0 = n_0^2$ adjacent neighbors. These macropixels span a size of $q = n_0(2\pi/W)$ in the momentum space and result in a ${\rm sinc}^2(qx/2){\rm sinc}^2(qy/2)$ illumination profile with diameter $D_{\rm{in}} = 3\pi/q$~\cite{2017_Hsu_nphys}. The corresponding macropixels transmission matrix $\tilde{\bf{t}}^{(\rm{macro})}$ is obtained by averaging over these $n_0^2$ columns in the original transmission matrix $\bf{t}$ with an appropriate normalization. The corresponding macropixels transmission eigenvalue $\tilde{\tau}^{(\rm{macro})}$ and eigenvalue distribution $p(\tilde{\tau}^{(\rm{macro})})$ are then numerically obtained from $\tilde{\bf{t}}^{(\rm{macro})\dag}\tilde{\bf{t}}^{(\rm{macro})}$. 

To get the theoretical prediction for this finite-area illumination, the equivalent fraction of controllable input modes is obtained through ${\rm{var}}(\tilde{\tau}^{(\rm{macro})})/{\rm{var}}(\tau)$~\cite{2013_Goetschy_PRL,2017_Hsu_nphys}, where ${\rm{var}}(\tau)$ represents the variance of $\tau$. We use this equivalent fraction of controllable modes to obtain the analytic prediction of the eigenvalue probability densities ${p\left( \tilde{\tau}^{(\rm{macro})} \right)}$ from FRM as shown in Fig. 3b of the main text.

\bibliography{suppl}